\documentclass[12pt]{article}
\usepackage[utf8]{inputenc}
\usepackage[margin=1in]{geometry}
\usepackage[dvipsnames]{xcolor}
\usepackage[colorlinks=true,linkcolor=Maroon,citecolor=BlueViolet, urlcolor=ForestGreen]{hyperref}
\usepackage[T1]{fontenc}
\usepackage{titling}
\usepackage{forest}
\usepackage{graphicx}
\usepackage{enumitem}
\usepackage{bookmark}
\usepackage{rotating} 
\usepackage{chngcntr} 
\usepackage{bm} 
\usepackage[flushleft]{threeparttable}
\usepackage{subcaption}
\usepackage{float, placeins}
\floatstyle{plain}
\usepackage{adjustbox}
\usepackage{threeparttable}
\usepackage{multirow}
\usepackage{dcolumn}
\usepackage{amsthm}
\usepackage{parskip}
\usepackage{thmtools}
\usepackage{thm-restate}
\usepackage{footnotebackref}
\usepackage{amssymb} 
\usepackage{amsmath}
\usepackage{booktabs} 
\usepackage{caption} 
\usepackage{bbm}
\usepackage{placeins}
\usepackage[toc,page,header]{appendix}
\usepackage{minitoc}
\usepackage{dsfont}
\usepackage[official]{eurosym}
\DeclareUnicodeCharacter{20AC}{\euro{}}

\usepackage{subfiles} 
\usepackage{xr}
\usepackage{setspace}
\usepackage[comma]{natbib}
\usepackage[linesnumbered, ruled]{algorithm2e}
\usepackage{cleveref} 

\usepackage{ragged2e}

\usepackage{tikz,float}
\usetikzlibrary{decorations} 
\usetikzlibrary{shapes,decorations,arrows,calc,arrows.meta,fit,positioning}
\tikzset{
    -Latex,auto,node distance =1 cm and 1 cm,semithick,
    state/.style ={ellipse, draw, minimum width = 0.7 cm},
    point/.style = {circle, draw, inner sep=0.04cm,fill,node contents={}},
    bidirected/.style={Latex-Latex,dashed},
    el/.style = {inner sep=2pt, align=left, sloped}
}

\usepackage{listofitems} 
\usetikzlibrary{arrows.meta} 
\usepackage[outline]{contour} 
\contourlength{1.4pt}
\tikzset{>=latex} 
\usepackage{xcolor}
\colorlet{myblue}{blue!80!black}
\colorlet{mydarkblue}{blue!40!black}
\colorlet{mydarkgreen}{green!30!black}
\tikzstyle{node}=[thick,circle,draw=myblue,minimum size=22,inner sep=0.5,outer sep=0.6]
\tikzstyle{node in}=[node,green!20!black,draw=teal!30!black,fill=teal!25]
\tikzstyle{node hidden}=[node,blue!20!black,draw=NavyBlue!30!black,fill=NavyBlue!20]
\tikzstyle{node out}=[node,red!20!black,draw=ForestGreen!30!black,fill=ForestGreen!20]
\tikzstyle{connect}=[thick] 
\tikzstyle{connect arrow}=[-{Latex[length=4,width=3.5]},thick,mydarkblue,shorten <=0.5,shorten >=1]
\tikzset{ 
  node 1/.style={node in},
  node 2/.style={node hidden},
  node 3/.style={node out},
}

\newcounter{assumption}
\renewcommand{\theassumption}{(A\arabic{assumption})}

\Crefname{assumption}{Assumption}{Assumptions}

\newcommand{\mycomment}[1]{}

\usepackage{etoolbox}

\appto\appendix{\counterwithin{equation}{section}}

\title{There's Nothing in the Air}

\author{Jacob Adenbaum \\ \small CUNEF Universidad 
  \and Fil Babalievsky \\ \small Census Bureau 
  \and William Jungerman\thanks{
    Adenbaum: CUNEF Universidad (\href{mailto:jacobadenbaum@gmail.com}{jacobadenbaum@gmail.com}),
Babalievsky: Census Bureau (\href{mailto:filipfba@gmail.com}{filipfba@gmail.com}),
Jungerman: University of North Carolina, Chapel Hill (\href{mailto:wjunger@unc.edu}{wjunger@unc.edu}). 
We thank Marlon Azinovic-Yang, Luca Flabbi, Kyle Herkenhoff, and Stan Rabinovich for helpful comments.
This work is generously supported by a public grant overseen by the French National Research Agency (ANR) as part of the ``Investissements d’Avenir'' program (reference: ANR-10-EQPX-17 - Centre d’accès sécurisé aux données – CASD).
Any opinions and conclusions expressed herein are those of the authors and do
not represent the views of the U.S. Census Bureau.
} \\ \small UNC, Chapel Hill }

\date{This version: \today} 

\usepackage{setspace}

\begin{document}
\renewcommand{\refname}{References}
\maketitle
\doparttoc 
\faketableofcontents 
\part{} 

\begin{abstract}
\noindent 
Why do wages grow faster in bigger cities? We use French administrative data to
decompose the urban wage growth premium and find that the answer has
surprisingly little to do with cities themselves. While we document substantially faster wage growth in larger cities, 80\% of the premium disappears after
controlling for the composition of firms and coworkers. 
We also document significantly higher
job-to-job transition rates in larger cities, suggesting workers climb the job
ladder faster. 
Most strikingly, when we focus on workers who remain in the same
job -- eliminating the job ladder mechanism -- the urban wage growth premium
falls by 94.1\% after accounting for firms and coworkers. The residual effect is
statistically indistinguishable from zero. These results challenge the view that
cities generate human capital spillovers ``in the air,'' suggesting instead that
urban wage dynamics reflect the sorting of firms and workers and the pace of job
mobility.

\vspace*{.4cm}
	
\vspace*{0.4cm}

\vspace*{0.1cm}
\end{abstract}

\clearpage


\newpage

\section*{Introduction}
\addcontentsline{toc}{section}{Introduction}
\newcommand{\redacted}{{\color{red}{X}}}

We have known, at least since \cite{smith1776wealth}, that the returns to labor differ across the city size distribution, and that these differences are deeply intertwined with the sorting of workers across space: 
\begin{quote}
    \it
    Industry, therefore, must be better rewarded, the wages of labour and the profits of stock must evidently be greater, in the one situation than in the other [town vs. country]. But stock and labour naturally seek the most advantageous employment. They naturally, therefore, resort as much as they can to the town, and desert the country.
\end{quote}

One of the central tasks of urban and regional economics has been to understand why wages are higher in bigger cities. Does it reflect spillovers that become stronger in dense cities, or does it reflect sorting of firms and workers? The answer to this question has important implications for the design of place-based policies and land-use regulations. 

Famously, \cite{marshall1890principles} argued that \textit{``the mysteries of the trade become no mysteries; but are as it were in the air''}, implying ambient learning spillovers from neighbors. This has become somewhat of a canonical explanation for why wages are higher in bigger cities. In this paper, we revisit the urban wage growth premium and provide new evidence that it is almost entirely mediated by the firms and coworkers that sort into cities.

In particular, we are motivated by a burgeoning literature on the role of coworkers in within-firm human capital accumulation \citep*{NixJMP2020,jaroschLearningCoworkers,GregoryJMP,herkenhoff2024production,adenbaum2024} which has found that a substantial fraction of on-the-job learning is attributed to having more-skilled coworkers. A natural implication of these findings is that, since bigger cities tend to have more productive firms and more skilled workers, sorting of firms and coworkers into bigger cities may be able to explain the dynamic urban wage premium documented in the seminal works of \cite{glaeser2001cities} and \cite{delaRocaPuga2017}. 

We focus on France and use rich matched employer-employee administrative data that allow us to track workers' labor markets trajectories over time to first document that the urban wage premium is significant: doubling the population of a commuting zone is associated with a 0.17\EUR{} increase in hourly wage growth. We then systematically decompose this premium into several components. First, we show that when we control for firm fixed effects, the dynamic urban wage premium falls by 61.5\%, suggesting that sorting of firms into cities accounts for a large share of the dynamic urban wage premium. Second, we show that adding detailed controls for coworkers, mirroring the functional forms in \citet*{jaroschLearningCoworkers} and \citet*{adenbaum2024}, without the firm fixed effects, similarly reduces the urban wage premium by 73.1\%. Adding both these coworker terms and the firm fixed effects explains 79.3\%. 

Next, we provide evidence that what remains of the premium is driven by job
mobility and labor market dynamics rather than human capital growth. We show
that job transitions occur more frequently in big cities: doubling city
population is associated with a 2\% increase in the rate of job transitions. We then redo our decomposition restricting the sample to workers that did not switch jobs, allowing us to control for these differences across cities. We find that controlling for firm fixed effects and coworkers now explains 94.1\% of the urban wage premium.

This is our main finding: adequately controlling for differences in firms, coworkers, and job mobility across cities almost entirely explains the urban wage growth premium. The residual effect after controlling for these three channels is both statistically and economically insignificant. Insofar as wages grow faster in bigger cities, our results suggest it arises from differences in distributions of firms and coworkers, as well as heterogeneity in job mobility patterns. These results shed new light on the underlying mechanisms behind the measured urban wage growth premium and leave little room for ambient spillovers ``\textit{in the air}''.

\paragraph{Related Literature.}

We contribute to the longstanding literature on the causes of the dynamic urban wage premium. This builds on the seminal work of \cite{glaeser2001cities}, who use survey data to find that a substantial part of the return to working in big cities accumulates over time and is retained when workers move across cities. In another influential contribution, \cite{delaRocaPuga2017} use Spanish administrative data to corroborate the findings of \cite{glaeser2001cities} on a larger scale. They too find that earnings growth is higher in bigger cities, and that workers retain these elevated earnings when they move across cities. \citet*{baum2012understanding} argue that faster wage growth in big cities is driven by greater returns to experience on the job. More recently, \citet*{eckert2022return} use the random allocation of refugees in Denmark to offer causal evidence that the returns to work are higher in bigger cities. \citet*{CardRothsteinYi2025} and \citet*{butts2023urban} likewise provide quasi-experimental evidence for the urban wage premium. Our contribution is to show that a large share of the urban wage premium appears to be mediated by differences in firms, coworkers, and job mobility across cities. This is not necessarily in tension with prior work that argues for a causal effect of cities on wages. Rather, we provide new evidence on the mechanisms behind this effect and argue that the mechanisms do not appear to be ambient spillovers but rather the concrete and measurable effects of firm and worker sorting, along with job mobility.

A large related literature attempts to understand the mechanism behind this dynamic urban wage premium. \cite{davisdingel19} introduce a dynamic multi-city model where faster growth in bigger cities is a result of costly idea exchange. \cite{martellini2022local} and \cite{CrewsJMP} both consider models of human capital growth in cities where spillovers depend on the skill distribution of the entire population.  \citet*{lindenlaubOhPeters2024} show that a steeper firm ladder in more productive cities can provide faster wage growth even without any human capital accumulation. Our goal in this paper is to provide suggestive evidence to help guide this theoretical literature, and to provide useful empirical targets for follow-on structural modeling.

Another set of papers are intermediate between the last two literatures, in that they attempt to provide suggestive reduced-form evidence on what mechanisms might be driving higher wages and faster wage growth in cities.  \citet*{CarryKleinmanNimier-David2025} use evidence from firm relocations to show that firms and coworkers seem to mediate a large share of the urban wage premium, leaving little room for the effect of the place itself on wages. This builds on the classic analysis of \citet*{combes2008spatial}, who show that sorting of workers by itself can explain a large share of regional income differences. \cite{Hong2024}, the paper most closely connected to our own, also finds evidence that faster wage growth in cities appears to be mediated by firms and coworkers. We contribute by providing new estimates from French administrative data, studying a richer set of coworker interactions, and providing new evidence on the role of the job ladder on what remains of the dynamic urban growth premium.

Our paper also contributes to the broader literature on human capital accumulation on the job. A burgeoning literature has studied human capital spillovers on the job and found evidence that workers learn from more-skilled coworkers. \citet*{jaroschLearningCoworkers}, \citet*{herkenhoff2024production}, and \citet*{adenbaum2024} study the sources of human capital growth using structural models and find a major role for learning from higher-skilled coworkers. \cite{NixJMP2020} finds that having more-educated coworkers causally increases wage growth. \cite{GregoryJMP} finds a role for the firms themselves in driving human capital accumulation. We contribute by showing that the mechanisms studied in these papers appear to account for a large share of the dynamic urban wage premium.

\ifSubfilesClassLoaded{%
    \bibliography{../../references}%
}{}

\end{document}

\newcommand{\redacted}{{\color{red}{X}}}

\section{Data}\label{Data}

We use administrative data from France made available to researchers by the French National Statistical Institute,
\textit{Institut national de la statistique et des études économiques} (INSEE). These data are widely used in labor and urban economics and allow us to match workers to their employers over time. Specifically, they are compiled from the \textit{Déclarations annuelles de données sociales} (DADS), mandatory tax forms firms in France must submit each year detailing information on their employees (such as wages, hours worked, occupation, etc.). Two main datasets are available. First, for the near-universe of workers in France, we have access to a \textit{short panel} that contains information on all firm-worker-year tuples but scrambles worker identifiers every two years. Second, we use \textit{long panel} which consists of,for a 1-in-12 sample of workers born in October, complete employment histories starting from the worker's first employment spell.

\paragraph{Sample construction.} We apply the same sample restrictions to both the short and the long panels, which we balance. We restrict the sample to private-sector employees aged 18-65 working in metropolitan France.\footnote{This excludes workers employed in overseas territories like Martinique and Guadeloupe.} When we use the short panel, we use the 2014-2015 cross-sections and when use the long panel, we use 1997 through 2019. We drop workers with missing information on key variables such wage, gender, and hours, as well as those with non-valid identifiers which we require for longitudinal analysis. We also drop workers from the sample earning less than 100\EUR{} per year, but we do not impose any upper limit on wages.

Investigating the role of coworker effects in explaining the urban wage growth premium means we also have to (1) define the boundaries of a city and (2) define the boundaries of a firm or a team, i.e. the set of coworkers. 

\begin{figure}[]
    \centering
    \includegraphics[width=0.7\linewidth]{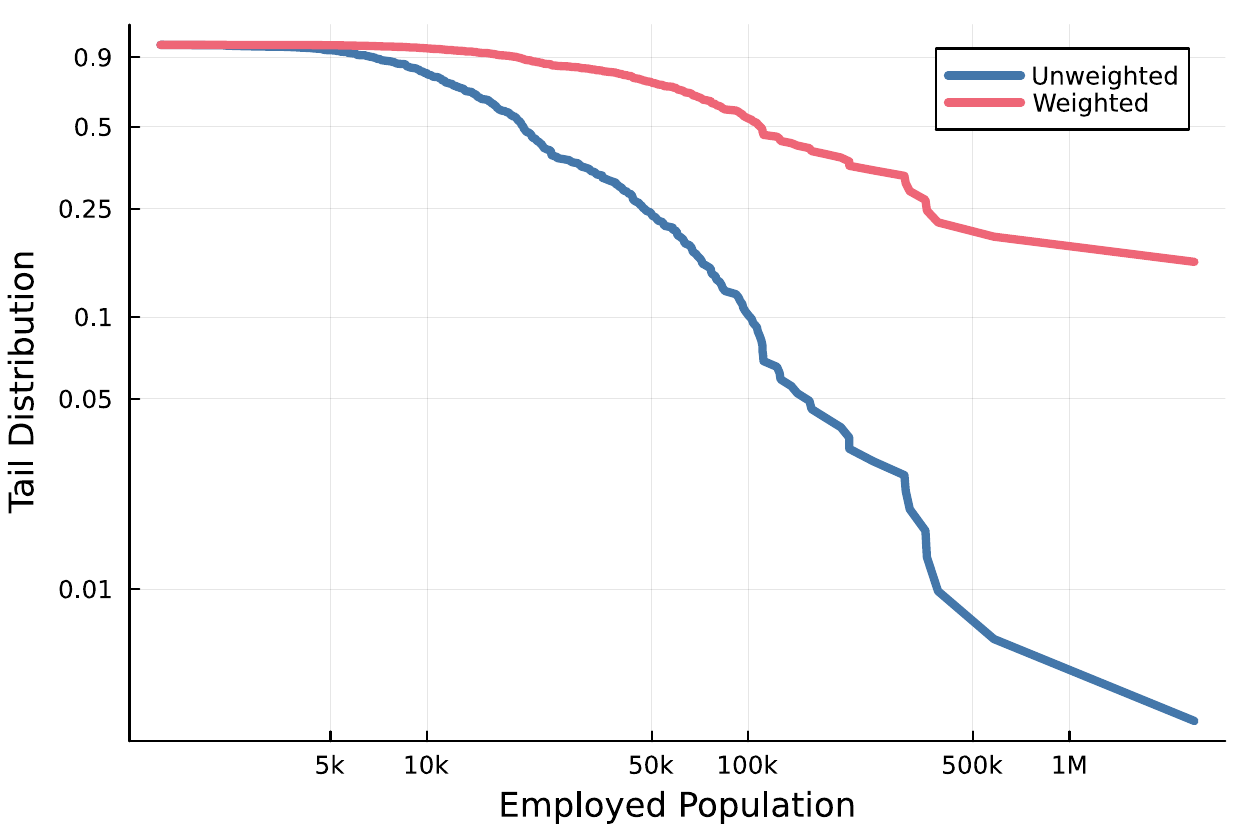}
    \caption{Pareto Tails of the Commuting Zone Size Distribution}
    \label{fig:CZ_size_distribution}
    \subcaption*{\footnotesize\textit{Note:} This figure plots the complementary cumulative distribution function (tail distribution),
    defined as the share of firms (establishment $\times$ 1-digit occupation) with more than $x$ employees, for both the unweighted and population-weighted firm size distributions.
    Computed in the 2015 DADS-Postes.}
 \end{figure}

\paragraph{Commuting zones.} Our notion of a city is a commuting zone (CZ), which is a set of municipalities that are economically connected. We use the 2010 version of the CZs, which are defined by INSEE and are based on the 1999 census. The CZs are constructed such that they capture the commuting patterns of workers, with the goal of identifying local labor markets. In \Cref{fig:CZ_size_distribution}, we plot the size distribution of CZs in France, measured by employed population. As expected, the long right tail is capturing the dominant role of Paris as the largest labor market in France. 

\paragraph{Defining a team.}
Following the approach in \citet*{adenbaum2024}, we identify 
a team as the intersection of an establishment and a 1-digit occupation code.
This definition is intended to capture the set of coworkers that 
a worker is likely to encounter in their day-to-day work environment.

The choice of team boundaries involves an inherent tradeoff.  
If defined too narrowly, it will exclude interactions 
between workers who actually collaborate and share knowledge.  
If defined too broadly, it will incorporate workers who never meaningfully 
interact.  
Our definition takes a conservative stance: we treat all workers at a given
establishment within the same 1-digit occupation as potential sources 
of interactions.  
We do not further partition teams by additional characteristics
like industry classification, sector, or echelon (e.g. entry-level versus manager), so that we can take as broad a stance as 
possible about the potential learning interactions between workers 
who may be interacting within an establishment.  

\Cref{table:occcodes} presents the 1-digit occupation categories we use, along 
with selected examples of the more detailed 2-digit and 4-digit codes that 
fall under \textit{Executives and High-Level Professionals}.  
If we chose to use finer measures of occupations for our classification 
of teams, this would imply that workers in closely related roles do not 
interact.  
For instance, using either 2 or 4-digit occupation codes would require us to
implicitly assume that dentists do not interact in meaningful ways with dental residents, or that lawyers do not interact with other legal professions.  
Such granular categorizations would artificially exclude what are likely 
productive learning relationships within the workplace. 

\Cref{table:selfflowrates} 
presents the transition rates across different 
organizational groupings in our data. 
Over the 2014 - 2015 period, we observe that 74.11\% of workers continue with the 
same team as we have defined it. 
Under our definition, there are three possible ways a worker can be classified 
as changing teams: relocating establishments, switching 1-digit occupations, 
or doing both at the same time.  
Transitions across establishments account for the bulk of the mobility we 
document, with approximately 20\% of workers moving between establishments 
year-over-year (not necessarily within the same firm).

The firm-level self flow rate is much higher than that for establishments, much 
as we would expect (the firm, after all, is a superset of the establishment).  The gap reflects workers who remain with the same employer 
but switch establishments \textit{within the same firm}.  
Because these sorts of transitions fundamentally change the composition 
of colleagues with whom a worker interacts on a day-to-day basis, 
we classify these individuals as having changed teams even though 
they remain at the same employer.\footnote{
  We abstract away from the question of whether or not workers are able to
  interact with their colleagues remotely, since our data window concludes 
  before the onset of the COVID-19 pandemic, and the corresponding widespread
  adoption of remote working arrangements.  
}

  \begin{table}
    \centering\begin{threeparttable}
        \caption{Self-Flow Rates}
        \label{table:selfflowrates}
        \centering
        \begin{tabular}{l c }
        \toprule 
        \toprule 
        & \textbf{Rate (\%)} \\
        OCC1 & 89.92 \\
        Firm & 83.64 \\
        Establishment & 79.16  \\ \\
        Establishment $\times$ OCC1 & 74.11 \\
        \bottomrule
        \bottomrule
        \end{tabular}
        \begin{tablenotes}[flushleft]\footnotesize
            \item \textit{Note:} 
            This table displays the share of workers who remain in the same organizational unit between 2014 and 2015, computed using the DADS-Postes data.
          \end{tablenotes}
    \end{threeparttable}
    \end{table}

In \Cref{table:team_summary_stats}, we provide summary statistics for our team definition, both unweighted and employment-weighted, as well as 
a comparison to alternative definitions. The unweighted average team size in our sample is 4.66 with a 5\% of teams being larger than 14 workers. However, the employment weighted mean is 152 with a long right tail. For instance, the 75th percentile is 83 and the 95th percentile is 535. Finally, we compute the mean number of 1-digit occupations per establishment, which is 3, suggesting that establishments on average have a very diverse set of workers.

\begin{table}
  \begin{adjustbox}{width=\textwidth}
  \centering\begin{threeparttable}
      \caption{Team Definition Summary Statistics}
      \label{table:team_summary_stats}
      \centering
      \begin{tabular}{l r r r r r r}
      \toprule 
      \toprule 
      & \textbf{Mean} & \textbf{SD} & \textbf{p25} & \textbf{Median} & \textbf{p75} & \textbf{p95}\\

      \textbf{Unweighted} \\
      Firm Size                                  & 8.30 & 202.06 & 1.00 & 1.00  & 4.00  & 20.00 \\ 
      Establishment Size                         & 7.15 & 43.96  & 1.00 & 1.00  & 4.00  & 24.00 \\ 
      Establishment $\times$ OCC1 Size           & 4.66 & 26.18  & 1.00 & 1.00  & 3.00  & 14.00  \\ \\

      \textbf{Employment-weighted} \\
      Firm Size                         & 4,925.21 & 17,113.05 & 12.00 & 96.00 & 930.00 & 34,005.00 \\ 
      Establishment Size                & 277.45   & 1,088.03  & 8.00  & 37.00 & 162.00 & 986.00 \\ 
      Establishment $\times$ OCC1 Size  & 151.72   & 646.95    & 4.00  & 17.00 & 83.00  & 535.00 \\ \\

      OCC1 per Establishment                & 3.06   & 1.26   & 2.00 & 3.00  & 4.00  & 5.00   \\
      
      \bottomrule
      \bottomrule
      \end{tabular}
      \begin{tablenotes}[flushleft]\footnotesize
          \item \textit{Note:} 
          Summary statistics of team size under our preferred
          definition (establishment interacted with 1-digit occupation) and
          alternative organizational boundaries, presented both with and without
          employment weights. Source: 2015 DADS-Postes. 
        \end{tablenotes}
  \end{threeparttable}
  \end{adjustbox}
  \end{table}

\newcommand{\redacted}{{\color{red}{X}}}

\FloatBarrier
\section{Decomposing the Urban Wage Premium}

We now turn to the main analysis. In a first stage, we regress wage growth on commuting zone fixed effects for workers who are employed in both periods of our short panel and stay in the same commuting zone. We estimate 
\begin{equation} 
  \label{eq:reg1}
  w_{i,t}=\nu w_{i,t-1} + \psi_{c(i,t)} +\epsilon_{i,t}
\end{equation}
where $w_{i,t}$ is the hourly wage of worker $i$ in commuting zone $c(i,t)$ at time $t$\footnote{Throughout the paper, we use commuting zone to mean the commuting zone of work. By construction of the commuting zones, residence commuting zone differs from work commuting zone for only a very small fraction of workers.}, and $\psi_{c(i,t)}$ is a commuting zone fixed effect. The coefficient $\nu$ captures the persistence of wages\footnote{Note that this specification allows us to test whether past wages matter for future wage growth. Subtracting by $w_{i,t-1}$ on both sides and rearranging yields 
$$ w_{i,t}-w_{i,t-1}=\psi_{c(i,t)}+(\nu-1) w_{i,t-1} +\epsilon_{i,t}$$
If $\nu = 1$, then wages follow a random walk. If $\nu <1$, then there is mean reversion. Finally, if $\nu >1$ then the initial advantages associated with being a high-wage worker compound over time.}, and the fixed effect $\psi_{c(i,t)}$ captures the static productivity differences across commuting zones.
Our main specifications will use hourly wages in levels, however, all of our results are robust to using logged hourly wages as the dependent variable.  
We report these results in \Cref{app:different_dv}.

In the second stage, we project down the estimated fixed effects $\hat \psi_{c(i,t)}$ from \Cref{eq:reg1} onto the log population $p_{c(i,t)}$ of commuting zone $c(i,t)$:

\begin{equation}\label{eq:FE_projection}
  \hat \psi_{c(i,t)} = \alpha_0 + \alpha p_{c(i,t)} + \epsilon_{i,t}
\end{equation}

By projecting the estimated fixed effects onto population in the second stage, we can quantify how much of the cross-commuting zone wage variation is systematically related to commuting zone size versus idiosyncratic commuting zone-specific factors. The coefficient $\alpha$ in the second stage tells us the semi-elasticity of hourly wages with respect to city population, while the $R^2$ reveals what fraction of the cross-city wage differences can be explained by population alone. As we progressively add controls in subsequent specifications, changes in both $\alpha$ and $R^2$ will reveal which mechanisms drive the urban wage premium. 

The results for our baseline specification are reported in Column (1) of \Cref{tab:decomposing_uwp_hourly}.  
This specification captures the raw urban wage growth premium that we see 
in the data, without any further controls, and we will refer back to it for the rest of the analysis as we seek to understand which channels and mechanisms drive the effect.   
In our second stage, we find a semi-elasticity of wages with respect to city size of 0.24.  
Taken at face value, this would suggest that doubling the population 
in a commuting zone is correlated with a 0.17€ increase in hourly wage growth, year over year. 
This is a relatively large effect, but it is almost identical to the results 
found in \cite*{delaRocaPuga2017} in Spain over a similar time period.\footnote{
\cite*{delaRocaPuga2017} report an \textit{elasticity} of earnings with
respect to population of 0.0241 after controlling for worker fixed effects (Column 4 of Table 1).  
Our results are not directly comparable, since we report semi-elasticities,
but we can do this as a back of the envelope calculation by using the fact
that in 2015 in Spain, average earnings were approximately 20,200\EUR{} \citep*{GRIDSpain} and assuming average
hours worked per week were 37.  This means that the average hourly wage was
10.49\EUR{} per hour.  This implies a corresponding semi-elasticity of 0.25,
which is almost identical to our estimate in Column (1). 
}
The $R^2$ of our second stage regression is relatively high: 
log population alone explains 39.3\% of the raw differences in average wage
growth across commuting zones.  

\begin{table}
  \centering
  \caption{Decomposing the Urban Wage Premium}
  \label{tab:decomposing_uwp_hourly}
  \resizebox{\linewidth}{!}{\begin{threeparttable}
    \begin{tabular}{l c c c c}
      \toprule 
      \toprule 
      & (1) & (2) & (3) & (4) \\
      \cmidrule(lr){2-5}
      & Baseline & + Lagged Firm FE & 
      Coworker Effects & + Lagged Firm FE \\
      \cmidrule(lr){1-5} \\
      \multicolumn{5}{l}{\textbf{Panel A: First Stage Regressions}} \\
      \\
      Lagged Wage ($\nu$) & 0.8072 & 0.6911 & 1.0188 & 0.9692 \\
                  & (0.0442) & (0.0644) & (0.0179) & (0.0415) \\
                        Higher-Wage Coworkers ($\tilde{\theta}^+_1$) & & & 0.1413 & 0.1324 \\
                                  & & & (0.0118) & (0.0480) \\
                        Lower-Wage Coworkers ($\tilde{\theta}^-_1$) & & & 0.1295 & 0.0779 \\
                                   & & & (0.0624) & (0.0876) \\
                        Nonlinear Effects ($\tilde{\theta}^{+}_2$) & & & -0.0000 & -0.0000 \\
                        & & & (0.0000) & (0.0000) \\
                        Nonlinear Effects ($\tilde{\theta}^{-}_2$) & & & -0.0001 & -0.0001 \\ 
                        & & & (0.0000) & (0.0000) \\ \\
                        Lagged Firm Fixed Effects & No & Yes & No & Yes \\
                        \\
                        Observations & 15,344,033 & 13,749,982 & 13,505,876 & 13,505,749 \\
                        $R^2$ & 0.1092 & 0.1318 & 0.1106 & 0.1370 \\
                        \\
                        \multicolumn{5}{l}{\textbf{Panel B: Second Stage Regressions}} \\
                        \\
                        Log Population ($\alpha$) & 0.23614 & 0.09086 & 0.06361 & 0.04877 \\
                               & (0.0278) & (0.0210) & (0.0131) & (0.0141) \\
                        \\
                        Observations & 304 & 304 & 304 & 304 \\
                        $R^2$ & 0.3938 & 0.1171 & 0.0921 & 0.0537 \\
                        \\
                        $\hat\alpha$ change relative to (1) & -- & -61.5\% & -73.1\% &  -79.3\%\\
                        $R^2$ change relative to (1) & -- & -70.3\% & -76.6\% & -86.4\% \\
      
      \bottomrule
      \bottomrule
    \end{tabular}
    \begin{tablenotes}
      \footnotesize
      \item \textit{Note:} Panel A reports first-stage regressions where the dependent variable is hourly wages. Column (1) shows the baseline specification from \Cref{eq:reg1} with commuting zone fixed effects and lagged wages. Column (2) adds firm fixed effects following \Cref{eq:reg2}. Column (3) includes the full coworker effects specification from \Cref{eq:reg4}. Column (4) combines coworker effects with firm fixed effects following \Cref{eq:reg5}. Panel B reports the projection of estimated fixed effects from the first stage onto logged commuting zone population following \Cref{eq:FE_projection}. Standard errors are clustered at the commuting zone level. Sample consists of all workers in the short panel with residualized wages for age and gender.
    \end{tablenotes}
  \end{threeparttable}}
\end{table}

Next, we add lagged firm fixed effects and estimate the equation
\begin{equation}
  \label{eq:reg2}
  w_{i,t}=\psi_{c(i,t)}+\mu_{j(i,t-1)}+\nu w_{i,t-1} +\epsilon_{i,t}
\end{equation}
where we define the firm $j(i,t-1)$ to be everyone with the same 1-digit
occupation code and the same employer.\footnote{
  Notice that our definition of a firm here differs from our definition of a
  team or set of coworkers described in \Cref{Data}. We need
  to use firm IDs and not establishment IDs here because we are already
  including commuting zone fixed effects. We cannot separately identify
  establishment times occupation codes fixed effects from commuting zone fixed
  effects.
}
The firm fixed effects capture any invariant characteristics of the firms which
may be relevant for wage growth at the worker level, including the industry, the
type of work they are engaged in, and other invariant characteristics of their
workforce composition.  For instance, if a firm in our sample consistently has a
better composition of coworkers, then this will be absorbed in the firm effect.
As before, in the second stage we project
down the estimated commuting zone fixed effects onto log population. The results
are reported in Column (2) of \Cref{tab:decomposing_uwp_hourly}.

Having added firm fixed effects, it is striking to examine our measure of the 
urban wage growth premium ($\alpha$).  We see that the semi-elasticity of hourly
wages with respect to commuting zone population falls from 0.24 to 0.09, a
drop of 61.5\%.  By adding firm fixed effects, we account for the fact that 
the composition of firms varies between big cities and small cities.  
In \Cref{tab:fe_pop} we regress the employment-weighted average firm fixed effect by commuting zone on log population.  
We see that high wage growth firms tend to be located in larger cities.  This
underlying pattern of firm sorting confounds the raw estimates of the urban wage
growth premium, and causes it to overstate the importance of population size per
se on the growth rate of wages.  We see as well that after controlling for firm
effects, population becomes dramatically less informative in explaining average
changes in wages across commuting zones, only accounting for 11.7\% of the
variation (compared to 39.3\% in our baseline case).  

\begin{table}[H]
  \centering
  \caption{Projecting Firm Fixed Effects on Commuting Zone Size}
  \label{tab:fe_pop}
  \begin{threeparttable}
    \begin{tabular}{l c}
      \toprule 
      \toprule 
      & (1) \\
      \cmidrule(lr){2-2}
      Dependent variable & Firm Fixed Effect \\
      \cmidrule(lr){1-2} \\
      
      Log Population & 0.0496 \\
            & (0.0121) \\\\
      
      Observations & 304 \\
      $R^2$ & 0.0530  \\
      
      \bottomrule
      \bottomrule
        \end{tabular}
        \begin{tablenotes}
      \footnotesize
      \item \textit{Note:} This table reports regressions of firm fixed effects on commuting zone size. 
        \end{tablenotes}
  \end{threeparttable}
\end{table}


\begin{figure}
  \begin{center}
    \includegraphics[width = \textwidth]{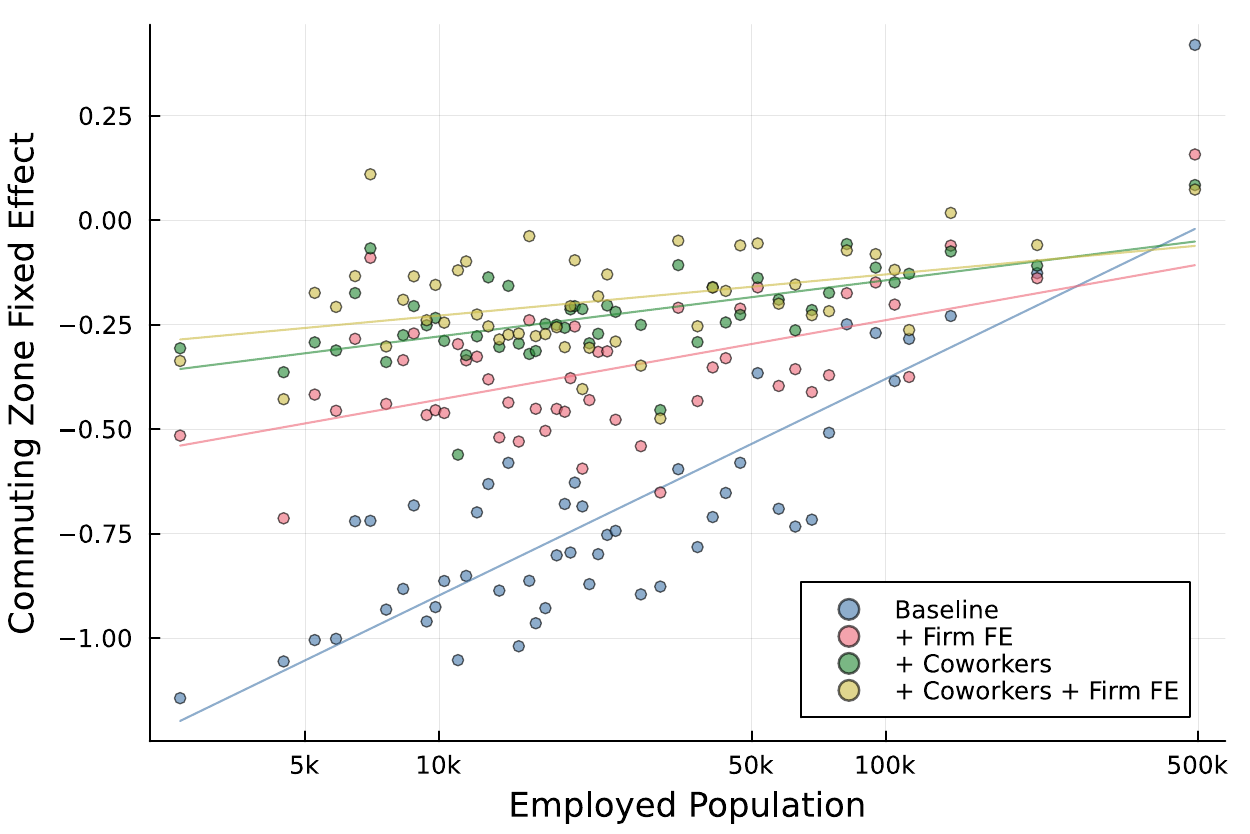}
  \end{center}
  \caption{City Fixed Effects vs Population}
  \label{fig:city_fes_vs_population}
  \subcaption*{
    \footnotesize
    \textit{Note:} 
    This figure plots the average commuting zone fixed effects 
    for each 2\% quantile bin of commuting zone log population.  
    ``Baseline'' plots the CZ fixed effects from \Cref{eq:reg1}.  
    ``Firm FE'' plots the CZ fixed effects from \Cref{eq:reg2}.
    ``Coworkers'' plots the CZ fixed effects from \Cref{eq:reg4}.
    ``Coworkers + Firm FE'' plots the CZ fixed effects from \Cref{eq:reg5}. 
  }
\end{figure}

While firm fixed effects can capture time-invariant firm characteristics,
recent evidence suggests that firm-specific features may not be the primary
driver of wage dynamics.  
\cite*{adenbaum2024} decompose the sources of human capital growth 
on the job, and find that peer effects account for more than 50\% of the variation in human capital growth for workers in France.  
Of particular importance is the effect of more skilled coworkers, which
contributes 52.5\% of the variation in human capital growth, even while
accounting for workplace and individual effects.
This suggests that who you work with matters more than where you work.
Moreover, unlike firm fixed effects, which capture invariant features that are
common across all their establishments, the composition of one's coworkers is
quintessentially local.  
If the composition of coworkers varies substantially across commuting zones, due
perhaps to the endogenous mobility choices of workers, small differences in the
growth rates of skills across cities, or firm sorting behavior, then we should
expect that the opportunities to learn from coworkers will vary with city size.  

To test this hypothesis, we augment our baseline specification with controls for
peer effects, as in \cite*{adenbaum2024} and \cite*{jaroschLearningCoworkers}.  
For each worker $i$ in our sample, we identify the set $\mathbb W_{i,t}^+$ of $i$'s coworkers who are paid more than $i$, and the set $\mathbb W_{i,t}^-$ of coworkers who are paid less than $i$.  
We compute over each of these sets the average deviation in wages from worker $i$'s wage, and the average squared deviation (which captures non-linearity in the learning function).\footnote{
  It is not obvious a priori how to compute these coworker controls in a computationally
  feasible way, since the naive algorithm's cost is quadratic in the size of the
  dataset.  In \Cref{app:quadratic_expansion} we show how to compute these
  coworker controls in linear time by carefully rewriting the problem using
  running sums.  
}
Consistent with our approach here, we identify the relevant set of coworkers in
the previous period $t-1$, since the wage growth we see between $t-1$ and $t$
should reflect human capital accumulation in the previous period, with the
coworkers at time $t-1$ rather the present day coworkers.  
Our choice to separately control for the effects of higher and lower paid workers 
reflects the findings in \cite*{herkenhoff2024production} that the 
human capital production function is nonlinear, with peer effects that 
depend on a coworker's relative position to the worker in question.  
We then estimate the equation
\begin{equation}
  \label{eq:reg4}
  \begin{aligned}
  w_{i, t} 
  &=  \psi_{c(i,t)} + \nu w_{i,t-1}
  + \underbrace{
    \tilde \theta^+_1 \sum_{j \in \mathbb W_{i,t}^+}  
          {w_{j,t-1}-w_{i,t-1} \over n_{k(i)} - 1}  
  }_\text{Higher-Wage Coworkers}
  \; + \;
  \underbrace{
    \tilde \theta^-_1 \sum_{j \in \mathbb W_{i,t}^-}  
          {w_{j,t-1}-w_{i,t-1} \over n_{k(i)} - 1} 
  }_\text{Lower-Wage Coworkers} \\
  & 
  + \underbrace{
      \tilde \theta^+_2 \sum_{j \in \mathbb W_{i,t}^+}  
            {(w_{j,t-1}-w_{i,t-1})^2 \over n_{k(i)} - 1}  
    \; + \;
      \tilde \theta^-_2 \sum_{j \in \mathbb W_{i,t}^-}  
            {(w_{j,t-1}-w_{i,t-1})^2 \over n_{k(i)} - 1} 
    }_\text{Nonlinear Effects}
    \; + \; \epsilon_{i,t} \\
  \end{aligned}
\end{equation}

Note that in this regression equation, we do not yet include firm fixed effects. 
We see that our measure of the urban wage growth premium 
declines even further in this specification, with the semi-elasticity of hourly wages with respect to commuting zone population 
falling from 0.24 in our baseline case to 0.06 with coworker
effects, a drop of 73.1\%.  
This is of a similar magnitude to the decline that we see in \cref{eq:reg2},
which we interpret as an indication that our coworker controls are capturing much of the same variation as the firm fixed effects.
The quality of a firm's workforce (and the corresponding potential for learning from one's coworkers) is common across all the workers at the firm.
However, our coworker controls allow us to capture the fact that 
individual workers within the firm may have different potential for future human
capital growth depending on their position within the within-firm human capital
distribution.  
For example, the most skilled worker in a firm has no one else to learn from, and our coworker controls can capture this.

In principle, we would like to account for both these sorts of coworker controls \textit{and} for invariant firm effects.  
To do so, we repeat the coworker specification from \cref{eq:reg4} adding in
lagged firm FE.
\begin{equation}
  \label{eq:reg5}
  \begin{aligned}
  w_{i, t} 
  &=  \psi_{c(i,t)} + \nu w_{i,t-1}+\alpha_{j(i,t-1)}
  + \underbrace{
    \tilde \theta^+_1 \sum_{j \in \mathbb W_{i,t}^+}  
          {w_{j,t-1}-w_{i,t-1} \over n_{k(i)} - 1}  
  }_\text{Higher-Wage Coworkers}
  \; + \;
  \underbrace{
    \tilde \theta^-_1 \sum_{j \in \mathbb W_{i,t}^-}  
          {w_{j,t-1}-w_{i,t-1} \over n_{k(i)} - 1} 
  }_\text{Lower-Wage Coworkers} \\
  & 
  + \underbrace{
      \tilde \theta^+_2 \sum_{j \in \mathbb W_{i,t}^+}  
            {(w_{j,t-1}-w_{i,t-1})^2 \over n_{k(i)} - 1}  
    \; + \;
      \tilde \theta^-_2 \sum_{j \in \mathbb W_{i,t}^-}  
            {(w_{j,t-1}-w_{i,t-1})^2 \over n_{k(i)} - 1} 
    }_\text{Nonlinear Effects}
    \; + \; \epsilon_{i,t} \\
  \end{aligned}
\end{equation}

It is worth taking a moment to discuss the question of separate identification
of the effects here.  In principle, we can separately identify coworker and
firm effects off of the average growth rate of wages by firm, and the gradient
of wage growth \textit{within the firm} against a worker's position within the
wage distribution of the firm.  
That is, if workers who are lower in the wage distribution consistently grow
faster than the firm average, then this will be attributed to $\tilde\theta_1^+$.
In practice, we expect that these effects are likely to be highly co-linear with the firm effects, and difficult to disentangle statistically.  
However, for our purposes this does not pose a substantial problem: we are not
interested in assigning any significant interpretation to the coefficients of our coworker controls.  
Rather, our object of interest is $\hat \alpha$: the correlation between the
commuting zone fixed effects $\psi_{c(i,t)}$ and the population of the
commuting zone $p_{c(i,t)}$, after partialling out the effects of our other
control variables.  

We can see the results from this specification in column (4) of \Cref{tab:decomposing_uwp_hourly}.  
We find that with \textit{both} coworker controls and lagged firm fixed effects,
the measured urban wage growth premium is somewhat smaller than with coworker
effects alone: the semi-elasticity drops to 0.048, a 79.3\% fall from our
baseline.
Another way to report these results is to plot the average commuting zone fixed effects 
for various quantile bins of log population, for each specification. 
We do this in \Cref{fig:city_fes_vs_population}, and we see that the slope of the line of best 
fit decreases dramatically when we add either firm fixed effects, or coworker controls.  
The commuting zone fixed effects look fairly similar controlling for either firm fixed effects or coworker fixed effects, suggesting that these two controls are capturing very similar variation 
in the underlying quality of the learning environments across firms. 
It is telling that when we compare the fixed effects controlling for coworkers (green) versus coworkers and firm fixed effects (gold), the results are almost indistinguishable to the eye.  
This reflects the fact that our measures of the urban wage growth premium ($\alpha$) are 
statistically indistinguishable between columns (3) and (4) in
\Cref{tab:decomposing_uwp_hourly}.

\paragraph{Robustness.} We relegate several robustness checks to the appendix. First, in \Cref{app:different_dv} we repeat the analysis for logged hourly wages. The results are reported in \Cref{tab:decomposing_uwp_log_hourly}, respectively. 
These results show extremely similar qualitative patterns to what we find in our
main specification. In fact, in log wages, the results are more striking, with a 98.0\% 
decline relative to our baseline specification when we restrict to job stayers and add coworker controls and firm fixed effects.  
Second, in \Cref{app:wage_growth}, we repeat the analysis using wage growth as the dependent variable in the first stage. For example, \Cref{eq:reg1} becomes 
\begin{equation} 
  \tag{1'}
  \label{eq:reg1'}
  \frac{w_{i,t} - w_{i,t-1}}{w_{i,t} + w_{i,t-1}} \cdot 2
  =  \psi_{c(i,t)} +\epsilon_{i,t}
\end{equation}
The results are reported in \Cref{tab:decomposing_uwp_wage_growth}.
In these specifications, we find that after accounting for coworker controls, or
coworkers controls and firm fixed effects, our measures of the urban wage growth
premium are indistinguishable from zero.
When we restrict attention only to job stayers, after controlling for coworker controls and firm fixed effects, the sign of the urban wage growth coefficient ($\alpha$) even becomes negative, although it is not statistically distinguishable from zero.  

Finally, in \Cref{app:other_moments}, we repeat the analysis using alternative 
measures of the within firm wage distribution to explore how much of our coworker controls can be explained by standard moments.   
In particular, we consider the mean, variance, skewness, kurtosis, and other
percentiles of the within firm wage distribution (the 1st, 10th, 90th, and 99th
percentiles).  Consistent with our previous specifications, where we allow the 
effect for a given worker to depend on their relative position within the wage
distribution, we introduce all of these controls for the distribution of 
coworkers and allow for interaction terms with the worker's wage.  The results
are reported in \Cref{tab:decomposing_uwp_moments}.  Even with all of these
controls for the distribution of workers within the firm, our measure of the
urban wage growth premium ($\alpha$) only falls by 42.5\% relative to our
baseline, compared to 73.1\% with our preferred coworker controls.  
A sensible choice of the specification for coworker controls, motivated by 
the underlying theory of human capital accumulation within the firm,
is key to explaining the variation that we see in the data.  

\paragraph{Discussion.}
These results suggest that insofar as we observe an urban wage growth
premium among workers in France, around 80\% of the underlying mechanism is
mediated through differences in the composition of both firms and coworkers
across urban areas.
This underlying mechanism in principle is a combination of several things:
First, workers could experience wage growth because their skills have increased
through learning.  
Second, workers' wages could increase because they have switched jobs, and 
this has allowed them to move up the wage ladder.  
Third, workers wages could have increased in their current job 
for various non-learning reasons (e.g, if an outside offer gives them leverage to bid up their current wages).  

Our learning controls and firm effects speak directly to the first mechanism.  
However, this approach to measuring the urban wage growth premium conflates the effects of learning with the effects of job to job transitions.
If workers transition jobs faster in larger cities, then mechanically we would
observe this as a correlation between the commuting zone fixed effects and
population.  
In the next section, we investigate whether or not this is the case: that is, do workers in bigger cities change jobs faster?

\subsection{Employment Transitions}

To investigate the question of whether or not the job transition rates 
differ across cities, in \Cref{eq:EE_1}, we run a simple regression of employer-to-employer (EE) transition rates as a function of log city population: 
\begin{equation}
  \label{eq:EE_1}
  EE_{i,t} = \alpha_0 + \alpha p_{c(i,t)} + \epsilon_{i,t}
\end{equation}
We use a cross-section from the short panel, which records a worker's employer
in the current period and in the lagged period. Note that in this case, our
notion of the employer is the same as we use for our definition of the 
set of coworkers: we identify a worker as experiencing a job transition if 
they change \textit{establishments or 1-digit occupations} 
from one year to the next.
This means that we treat both internal movements between establishments within a
firm, or changes in roles large enough to change the worker's 1-digit occupation code, 
as \textit{job transitions}.  
For our purposes, such changes in the worker's team within the firm are very
similar to external moves, in terms of the expected effect they will have on
wages.  
We report the results in the first column of Table \ref{tab:EE_mechanisms}, and
find a significant and positive relationship between city size and EE transition
rates.

Next, in Equation \ref{eq:EE_2} we add a control for lagged wage, i.e, the wage in the first period of the two-period short panel:
\begin{equation}
  \label{eq:EE_2}
  EE_{i,t} = \alpha_0 + \alpha p_{c(i,t)} + \beta w_{i,t-1}+\epsilon_{i,t}
\end{equation}
This regression provides evidence on whether workers leave a job of a given quality more or less quickly. This may be driven by better matching in thicker labor markets, or by the higher prevalence of better jobs in bigger cities that can win poaching fights more easily. We report the results in the second column of Table \ref{tab:EE_mechanisms} and again find a significant and positive city size effect.
As before, in \Cref{tab:EE_mechanisms_log} we repeat the analysis for logged hourly wages and find similar results.

\begin{table}[H]
  \centering
  \caption{Job-to-Job Transition Rates and City Size}
  \label{tab:EE_mechanisms}
  \begin{threeparttable}
    \begin{tabular}{l c c}
      \toprule 
      \toprule 
      & (1) & (2) \\
      \cmidrule(lr){2-3}
      Dependent variable & EE Transition & EE Transition \\
      \cmidrule(lr){1-3} \\
      
      Log Population & 0.0296 & 0.0368 \\
                    & (0.0017) & (0.0012) \\
      Lagged Hourly Wage & & -0.0018 \\
                    & & (0.0005) \\\\
      
      Controls & No & Lagged Wage \\
      Fixed Effects & No & No \\
      SE Clustering & CZ & CZ \\ \\
      
      Observations & 15,344,033 & 15,344,033 \\
      $R^2$ & 0.0023 & 0.0062  \\
      
      \bottomrule
      \bottomrule
    \end{tabular}
    \begin{tablenotes}
      \footnotesize
      \item \textit{Note:} This table reports regressions of job-to-job (EE) transition rates on city size. Column (1) shows the baseline specification from \Cref{eq:EE_1}. Column (2) adds lagged hourly wages as a control following \Cref{eq:EE_2}. Standard errors are clustered at the commuting zone (CZ) level. The sample consists of all workers from one cross-section of the short panel.
    \end{tablenotes}
  \end{threeparttable}
\end{table}

These results stand in contrast to \cite{martellini2022local} and \cite{Hong2024}, who find little evidence for faster job transitions in bigger cities. Understanding the source of this discrepancy is an important goal for future work.
It is also important to note that, while the results support the idea that the
unexplained remainder of the dynamic urban wage premium might be explained by
differences in the rate of EE moves, this is not necessarily evidence of
increasing returns to search. As \citet*{lindenlaubOhPeters2024} show, the returns
to EE transitions may be higher because of positive firm sorting towards more
productive cities. Hence, while we provide evidence that labor market
transitions can help explain what is left of the dynamic urban wage premium
after we control for firms and coworkers, more work is still required to
understand exactly what drives this faster climb up the job ladder in bigger
cities. 

\subsection{Job stayers}
Motivated by our evidence from the previous section that larger cities in France 
appear to have a higher job-to-job transition rate, 
we now restrict our attention to the subsample of workers
who have not transitioned jobs in the past year.  
This restriction allows us to separate out the effect of faster transitions from faster 
skill growth because job stayers, by definition, cannot have moved up on the job
ladder within the previous year.   
Changes in their wages must reflect either learning or bargaining.  

\begin{figure}
  \begin{center}
    \includegraphics[width = \textwidth]{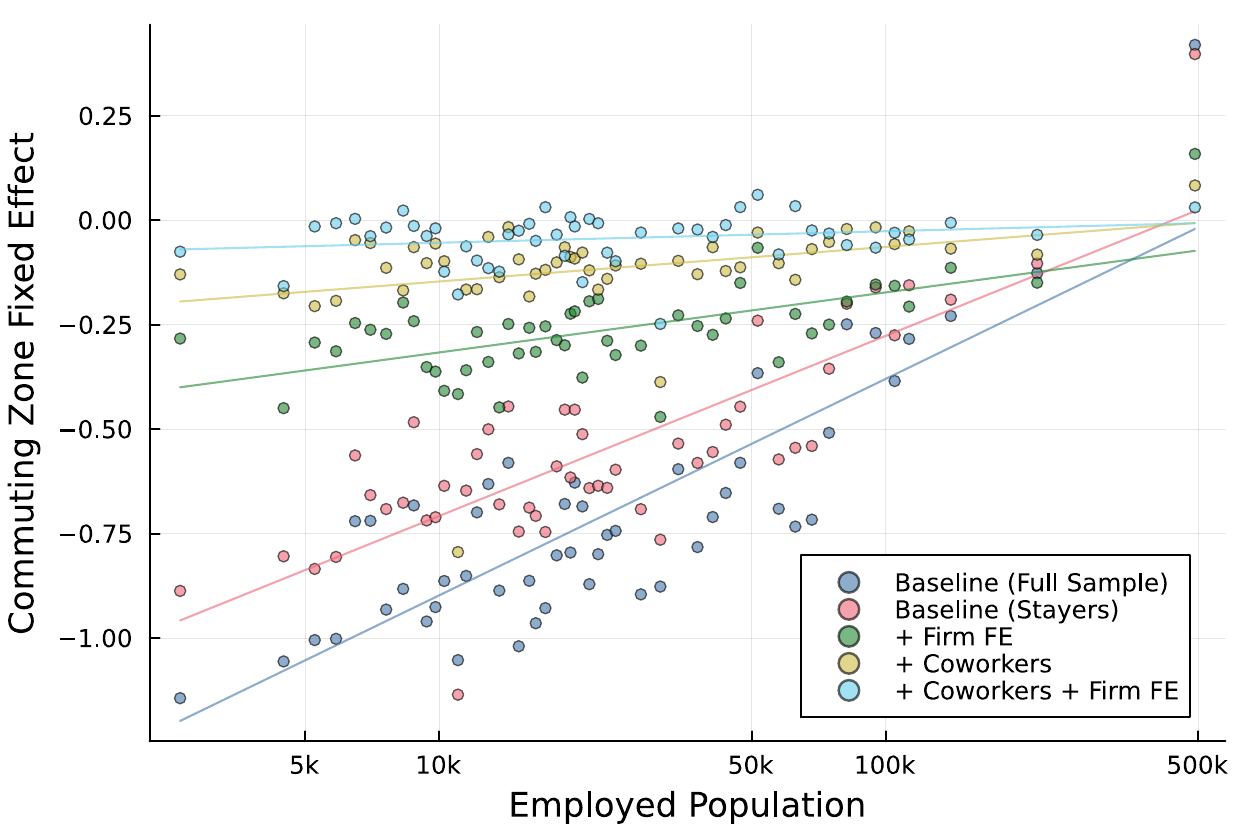}
  \end{center}
  \caption{City Fixed Effects vs Population for Job Stayers}
  \label{fig:city_fes_vs_population_stayers}
  \subcaption*{
    \footnotesize
    \textit{Note:} 
    This figure plots the average commuting zone fixed effects 
    for each 2\% quantile bin of commuting zone log population, in 
    our sample restricted to job stayers.    
    ``Baseline (Full Sample)'' plots the CZ fixed effects from \Cref{eq:reg1} for the full sample.  
    ``Baseline (Stayers)'' plots the CZ fixed effects from \Cref{eq:reg1} when we restrict to job stayers.  
    ``Firm FE'' plots the CZ fixed effects from \Cref{eq:reg2} for job stayers.
    ``Coworkers'' plots the CZ fixed effects from \Cref{eq:reg4} for job stayers.
    ``Coworkers + Firm FE'' plots the CZ fixed effects from \Cref{eq:reg5} for job stayers. 
  }
\end{figure}

We re-run specifications \Cref{eq:reg1,eq:reg2,eq:reg4,eq:reg5} on this
subsample and report the results in \Cref{tab:decomposing_uwp_stayers}.
Relative to our full sample, we see a similar baseline urban wage growth premium with a measured semi-elasticity of 0.196 relative to 0.236 in the full sample. 
Adding firm fixed effects, the growth premium falls to 0.068. 
Controlling for coworker effects it falls to 0.040, and controlling for both it falls to 0.014.  
This constitutes a remarkable 94.1\% drop in the magnitude of the measured urban wage growth premium among job stayers.   
Moreover, the coefficient is statistically indistinguishable from zero at the 1\% level. 
In \Cref{fig:city_fes_vs_population_stayers} we plot the commuting zone fixed effects against log population. 
As before, we check that these results are robust to a variety of other specifications, reported in \Cref{app:different_dv}.  

So what exactly is ``in the air'' that makes wages grow faster in bigger cities?
Our results provide a surprisingly concrete answer:
when we focus on workers who remain in the same job, thereby eliminating the 
job ladder mechanisms, the urban wage growth premium falls by 94.1\% once 
we account for firms and coworkers.  
The residual semi-elasticity is not only statistically indistinguishable from zero
but also economically negligible.  
This decomposition suggests that the well-documented urban wage growth premium 
operates through three specific channels:
the quality of firms that locate in cities, the composition of coworkers within those 
firms, and the speed at which workers can climb the job ladder.  

This decomposition helps clarify the mechanisms underlying a phenomenon that 
has been central to urban economics since \cite{marshall1890principles}.  
Rather than diffuse knowledge spillovers that benefit all workers in a city,
our findings suggest that faster wage growth is remarkably localized.  
It depends on the specific firm you work for and the specific people you work with. 
For job stayers, once we know their firm and coworkers, 
knowing that they work in Paris versus a small town adds virtually no additional 
information about their wage growth prospects.
This aligns with recent structural evidence on 
the importance of learning interactions \textit{within the firm}, as in \cite*{adenbaum2024}, \cite*{herkenhoff2024production}, and \cite*{jaroschLearningCoworkers}.
In all of these frameworks, human capital accumulation operates through direct
interactions with specific coworkers rather than through ambient effects.  

The three channels we identify (firm quality, coworker composition, and job
mobility) together substantially account for the observed urban wage growth premium.
Our evidence on sorting shows that higher-quality firms concentrate in larger 
cities, creating a superior learning environment \textit{within the firm}.  
Meanwhile, our finding of higher EE job transition rates in larger cities 
suggests that the job ladder operates more quickly there, whether due to thicker 
labor markets that facilitate better matches or a steeper firm quality distribution 
as in \cite*{lindenlaubOhPeters2024}.  
What's notable is how cleanly these mechanisms decompose: for job stayers, the firm and coworker channels explain virtually everything, while job mobility explains the remainder.  
Researchers who are interested in modeling the dynamics of wage growth across
space need not resort to ambient spillovers operating outside these channels, 
which are difficult to interpret as structural and microfounded mechanisms.

These findings offer a clearer picture of how urban wage dynamics 
actually work.  By decomposing the urban premium into its constituent parts, we
can see that the advantages of cities for wage growth are remarkably concrete 
and measurable.  
For workers who remain in the same job, city size itself appears to play no direct
role in wage growth once we account for who they work with.  
While understanding the deeper mechanisms that drive firms and workers 
to sort into cities, and exactly what generates faster job transitions there, remains an important area for future research, our decomposition provides 
sharp empirical targets for such work.  
The mechanisms behind urban wage growth turn out to be more immediate and tangible 
than the metaphor of knowledge ``in the air'' might suggest. 
They operate through the specific people you work with 
and the specific opportunities available to change jobs.
In this fundamental sense, there really is nothing in the air.  

\begin{table}
  \centering
  \caption{The Urban Wage Premium Holding Job Mobility Constant}
  \label{tab:decomposing_uwp_stayers}
  \resizebox{\linewidth}{!}{\begin{threeparttable}
    \begin{tabular}{l c c c c c}
      \toprule
      \toprule
      & (1) & (2) & (3) & (4) & (5) \\
      \cmidrule(lr){2-6}
      & Baseline (Full Sample) & Baseline (Stayers) & + Lagged Firm FE & 
      Coworker Effects & + Lagged Firm FE \\
      \cmidrule(lr){1-6} \\
      \multicolumn{6}{l}{\textbf{Panel A: First Stage Regressions}} \\
      \\
      Lagged Wage ($\nu$) & 0.8072 & 0.8610 & 0.7461 & 0.9856 & 0.8728 \\
                   & (0.0442) & (0.0636) & (0.0973) & (0.0209) & (0.0573) \\
          Higher-Wage Coworkers ($\tilde{\theta}^+_1$) & & & & 0.1389 & 0.1634 \\
                                & & & & (0.0156) & (0.0542) \\
          Lower-Wage Coworkers ($\tilde{\theta}^-_1$) & & & & -0.0408 & -0.1863 \\
                                 & & & & (0.0639) & (0.1125) \\
          Nonlinear Effects ($\tilde{\theta}^{+}_2$) & & & & -0.0000 & -0.0000 \\
          & & & & (0.0000) & (0.0000) \\
          Nonlinear Effects ($\tilde{\theta}^{-}_2$) & & & & -0.0001 & -0.0001 \\
          & & & & (0.0000) & (0.0000) \\ \\
          Lagged Firm Fixed Effects & No & No & Yes & No & Yes \\
          \\
          Observations & 15,344,033 & 9,175,509 & 8,139,602 & 8,195,025 & 8,008,451 \\
          $R^2$ & 0.1092 & 0.6849 & 0.7298 & 0.7351 & 0.7710 \\
          \\
          \multicolumn{6}{l}{\textbf{Panel B: Second Stage Regressions}} \\
          \\
          Log Population ($\alpha$) & 0.23614 & 0.19632 & 0.06798 & 0.03954 & 0.01405 \\
                       & (0.0278) & (0.0230) & (0.0162) & (0.0132) & (0.0068) \\
                \\
                Observations & 304 & 304 & 304 & 304 & 304 \\
                    $R^2$ & 0.3938 & 0.2923 & 0.1253 & 0.0161 & 0.0104 \\
                    \\
                  $\hat\alpha$ change relative to (1) & -- & -16.9\% & -71.2\% & -83.3\% & -94.1\% \\
                  $R^2$ change relative to (1) & -- & -25.8\% & -68.2\% & -95.9\% & -97.4\% \\
      
      \bottomrule
      \bottomrule
    \end{tabular}
    \begin{tablenotes}
      \footnotesize
      \item \textit{Note:} Panel A reports first-stage regressions where the dependent variable is hourly wages. Column (1) shows the baseline specification from \Cref{eq:reg1} for the full sample. Columns (2)-(5) restrict to job stayers. Column (2) shows the baseline specification from \Cref{eq:reg1} restricting the sample to job stayers. Column (3) adds firm fixed effects following \Cref{eq:reg2}. Column (4) includes the full coworker effects specification from \Cref{eq:reg4}. Column (5) combines coworker effects with firm fixed effects following \Cref{eq:reg5}. Panel B reports the projection of estimated fixed effects from the first stage onto logged commuting zone population following \Cref{eq:FE_projection}. Standard errors are clustered at the commuting zone level. Wages are residualized for age and gender.
    \end{tablenotes}
  \end{threeparttable}}
\end{table}

\FloatBarrier
\section*{Conclusion}\label{section:conclusion}
\addcontentsline{toc}{section}{Conclusion}
\newcommand{\redacted}{{\color{red}{X}}}

In this paper, we have used French administrative data to provide new evidence on the causes of the dynamic urban wage premium. We first demonstrated that this dynamic premium is substantially attenuated after adding firm fixed effects, suggesting that part of this premium is driven by the firms that choose to sort themselves into bigger cities. Next, we showed that the premium shrinks even further when adding controls for coworker wages, which suggests a role for learning from the higher human capital workers that sort themselves into cities. Crucially, we showed that moments of the worker wage distribution are not sufficient in themselves---what matters is the mass of coworkers with higher and lower wages than the reference worker. Combining firm and coworker controls greatly attenuated the dynamic urban wage premium in our full sample, and eliminated it altogether in our subsample of job stayers. Motivated by this discrepancy between job stayers and the full sample, we showed that employer-to-employer transition rates are higher in bigger French cities. This suggests that labor market factors (whether they are increasing returns to search or a steeper firm quality ladder) help explain the remaining part of the dynamic urban wage premium that survived our worker and firm controls. Taken at face value, our results leave little room for the canonical explanation for the dynamic urban wage premium, namely human capital spillovers ``\textit{in the air}''.

\ifSubfilesClassLoaded{%
    \bibliography{../../references}%
}{}

\end{document}

\newpage
\bibliography{../references}

\newpage

\newpage\appendix
\addcontentsline{toc}{section}{Appendix} 
\part{Appendix} 
\parttoc 
\newpage

\counterwithin{figure}{section}
\counterwithin{table}{section}
\renewcommand{\refname}{Appendix References}
\newcommand{\redacted}{{\color{red}{X}}}


\section{Data Appendix}
\subsection{Variable List}\label{app:variable_list}
\begin{itemize}
    \item[--] Hourly Wages: We use hourly wages to control for 
    variation in hours worked. Wages are \texttt{S\_BRUT} in the short panel and \texttt{SB} in the long panel.
    These are gross annual wages which include overtime and bonuses. We divide these by \texttt{NBHEUR}, the number of hours worked in the year, to get hourly wages.
    Finally, we use the CPI provided by INSEE to deflate all wages to 2015 euros.
    \item[--] Establishment: We use the INSEE provided establishment codes \texttt{SIRET} to define the boundaries of a team.
    \item[--] Occupation: We also use 1-digit occupation codes, the first digit of \texttt{PCS4}, to define a team.
\end{itemize}

\newpage 
\begin{table}[ht!]
    \centering
    \begin{threeparttable}
    \caption{Occupation Coding in DADS Data}
    \label{table:occcodes}
    \centering
    {\footnotesize
    \begin{tabular}{l*{4}{l}}
    \toprule
     \textbf{1} & & & \textbf{Farmers} \\ 
      \textbf{2} & & & \textbf{Craftsmen, Tradespeople, and Business Owners} \\ 
      \textbf{3} & & & \textbf{Executives and High-Level Professionals} \\ \\
      & 31 & & \textit{Independent Professionals} \\ \\
        & & 311c & Dentists \\
        & & 311d & Psychologists and Therapists \\
        & & 311e & Veterinarians \\
        & & 3121 & Lawyers \\ \\
        & 33 & & \textit{Public Sector Executives} \\  
        & 34 & & \textit{Professors, Scientific Professionals} \\ \\
        & & 342b  & Research Professors \\
        & & 344a  & Hospital Doctors Without an Independent Practice \\
        & & 344c  & Residents in Medicine, Dentistry and Pharmacy  \\ 
        & & 344d  & Salaried Pharmacists \\ \\
        & 35 & & \textit{Careers in Media, Arts, and Entertainment} \\ \\
        & & 352a & Journalists \\ 
        & & 352b & Literary Authors, Screenwriters  \\ 
        & & 353a & Newspaper Editors, Media Executives, Publishing Directors  \\ \\ 
        & 37 & & \textit{Corporate Administrative and Commercial Managers} \\ \\
        & & 372e & Legal Professionals  \\ 
        & & 375a & Advertising Executives  \\
        & & 376a & Financial Market Executives  \\ \\
        & 38 & & \textit{Engineers and Technical Managers} \\ \\
        & & 382b & Salaried Architects  \\
        & & 384b & Mechanical Manufacturing Engineers and Metalworking Managers  \\
        & & 387c & Production Process Engineers and Managers  \\ 
        & & 387d & Quality Control Engineers and Managers  \\ \\
      \textbf{4} & & & \textbf{Intermediate Professions} \\ 
      \textbf{5} & & & \textbf{Clerical Workers} \\ 
      \textbf{6} & & & \textbf{Manual Laborers} \\ 
      \textbf{9} & & & \textbf{Non-Coded} \\ 
    \bottomrule
    \end{tabular}
    }
    \begin{tablenotes}
        \footnotesize
        \item \textit{Note:} This table illustrates occupation coding in the DADS. There 
        are 6 distinct 1-digit occupation codes. To illustrate the granularity of 2-digit occupation coding, we also report a selection of the 2-digit and 4-digit codes for a representative example.
        The full list of occupation codes can be found on INSEE's website at \url{https://www.insee.fr/fr/information/2497952}.
    \end{tablenotes}
  \end{threeparttable}
  \end{table}


\subsection{Quadratic Expansion}\label{app:quadratic_expansion}

In this appendix, we derive computationally efficient formulas for the positive and negative components of the coworker effects in both the linear and quadratic cases, which we use to estimate \Cref{eq:reg4,eq:reg5}. These formulas can be computed in linear time after sorting the data, and do not require any look-aheads or look-backs through the data. We begin with the linear case. Suppose we have a sequence $\{x_i\}_{i=1}^N$ and we want to compute, for each $j\in \{1, 2, \dots, N\}$
\begin{align*}
y_j^+ := \frac{1}{N} \sum_{i=1}^N  \max\{x_i - x_j, 0\} 
\end{align*}
Without loss of generality, we will assume throughout that 
the sequence $\{x_i\}$ is sorted in ascending order, i.e. $x_1 \leq x_2 \leq x_3 \leq \cdots \leq x_N$. Moreover, we use $Z_j$ to denote the running sum 
$$Z_j := \sum_{i=1}^j x_i$$
which can conveniently be computed in a single pass through the data (i.e in linear time) since $Z_{j+1} = Z_j + x_{j+1}$.
Using this, we can rewrite $y_j^+$ as
\begin{align*}
y_j^+ &= \frac{1}{N} \sum_{i=j+1}^N (x_i - x_j) && (x_i \geq x_j \text{ for } i \geq j) \\
&= \frac{1}{N} \left[ \sum_{i=j+1}^N x_i - (N-j) x_j \right] \\
&= \frac{1}{N} \left[ Z_N - Z_j - (N-j) x_j \right]
\end{align*}
We can also do the same for the negative component
\begin{align*}
  y_j^- &:= \frac{1}{N} \sum_{i=1}^N \min\{x_i - x_j, 0\} \\
&= \frac{1}{N} \sum_{i=1}^{j-1} (x_i - x_j) && (x_i \geq x_j \text{ for } i \geq j)\\
&= \frac{1}{N} \left[ Z_{j-1} - (j-1) x_j \right] \\
&= \frac{1}{N} \left[ Z_j - x_j - (j-1) x_j \right] \\
&= \frac{1}{N} (Z_j - j x_j)
\end{align*}
We now turn to the quadratic terms and similarly define 
$$Z_{2,j} := \sum_{i=1}^j x_i^2$$ 
to be the running sum of the $x_i^2$ terms up to $j$. The positive component in the quadratic case can then be rewritten as 
\begin{align*}
y_{2,j}^+ &= \frac{1}{N} \sum_{i=1}^N \max\{x_i - x_j, 0\}^2 \\
      &= \frac{1}{N} \sum_{i=j+1}^N (x_i - x_j)^2 \\
      &= \frac{1}{N} \sum_{i=j+1}^N (x_i^2 - 2x_i x_j + x_j^2) \\
      &= \frac{1}{N} \left[ \sum_{i=j+1}^N x_i^2 - 2x_j \sum_{i=j+1}^N x_i + (N-j) x_j^2 \right] \\
      &= \frac{1}{N} \left[ (Z_{2,N} - Z_{2,j}) - 2x_j (Z_N - Z_j) + (N-j) x_j^2 \right]
\end{align*}
Similarly, for the negative component, we have
\begin{align*}
y_{2,j}^- &= \frac{1}{N} \sum_{i=1}^N \min\{x_i - x_j, 0\}^2 \\
      &= \frac{1}{N} \sum_{i=1}^{j-1} (x_i - x_j)^2 \\
      &= \frac{1}{N} \sum_{i=1}^{j-1} (x_i^2 - 2x_i x_j + x_j^2) \\
      &= \frac{1}{N} \left[ \sum_{i=1}^{j-1} x_i^2 - 2x_j \sum_{i=1}^{j-1} x_i + (j-1) x_j^2 \right] \\
      &= \frac{1}{N} \left[ Z_{2,j-1} - 2x_j Z_{j-1} + (j-1) x_j^2 \right] \\
      &= \frac{1}{N} \left[ (Z_{2,j} - x_j^2) - 2x_j (Z_j - x_j) + (j-1) x_j^2 \right] \\
      &= \frac{1}{N} \left[ Z_{2,j} - 2x_j (Z_j - x_j) + (j-2) x_j^2 \right] 
\end{align*}

\section{Results Appendix}

\subsection{Robustness to Log Wages}\label{app:different_dv}
\begin{table}[H]
  \centering
  \caption{Decomposing the Urban Wage Premium: Log Hourly Wages}
  \label{tab:decomposing_uwp_log_hourly}
  \resizebox{\linewidth}{!}{\begin{threeparttable}
    \begin{tabular}{l c c c c}
      \toprule 
      \toprule 
      & (1) & (2) & (3) & (4) \\
      \cmidrule(lr){2-5}
      & Baseline & + Lagged Firm FE & 
      Coworker Effects & + Lagged Firm FE \\
      \cmidrule(lr){1-5} \\
      \multicolumn{5}{l}{\textbf{Panel A: First Stage Regressions}} \\
      \\
      Lagged Wage ($\nu$) & 0.8313 & 0.5590 & 0.9141 & 0.7706 \\
                         & (0.0005) & (0.0010) & (0.0003) & (0.0020) \\
      Higher-Wage Coworkers ($\tilde{\theta}^+_1$) & & & 0.2299 & 0.0724 \\
                                                    & & & (0.0018) & (0.0027)\\
      Lower-Wage Coworkers ($\tilde{\theta}^-_1$) & & & 0.0315 &  -0.0357\\
                                                   & & & (0.0013) & (0.0032) \\
      Nonlinear Effects ($\tilde{\theta}^{+}_2$) & & & 0.0770 & 0.0784 \\
      & & & (0.0009) & (0.0010) \\
      Nonlinear Effects ($\tilde{\theta}^{-}_2$) & & & -0.0064 & -0.0240 \\ 
      & & & (0.0008) &   (0.0001) \\ \\
      Lagged Firm Fixed Effects & No & Yes & No & Yes \\
      \\
      Observations & 15,344,033 & 13,749,982 & 13,505,876 & 13,505,749 \\
      $R^2$ & 0.7531 & 0.7667 & 0.7621 & 0.7940 \\
      \\
      \multicolumn{5}{l}{\textbf{Panel B: Second Stage Regressions}} \\
      \\
      Log Population ($\alpha$) & 0.00813 & 0.00351 & 0.00327 & 0.00213 \\
                       & (0.0008) & (0.0008) & (0.0004) & (0.0007)\\
          \\
          Observations & 304 & 304 & 304 & 304 \\
          $R^2$ & 0.3756 & 0.0866 & 0.1908 & 0.0457 \\
          \\
          $\hat\alpha$ change relative to (1) & -- & -56.8\% & -59.9\% & -73.8\% \\
          $R^2$ change relative to (1) & -- & -76.9\% & -49.2\% & -87.8\% \\
      
      \bottomrule
      \bottomrule
    \end{tabular}
    \begin{tablenotes}
      \footnotesize
      \item \textit{Note:}  Panel A reports first-stage regressions where the dependent variable is log hourly wages. Column (1) shows the baseline specification from \Cref{eq:reg1} with commuting zone fixed effects and lagged wages. Column (2) adds firm fixed effects following \Cref{eq:reg2}. Column (3) includes the full coworker effects specification from \Cref{eq:reg4}. Column (4) combines coworker effects with firm fixed effects following \Cref{eq:reg5}. Panel B reports the projection of estimated fixed effects from the first stage onto logged commuting zone population following \Cref{eq:FE_projection}. Standard errors are clustered at the commuting zone level. Sample consists of all workers in the short panel with residualized wages for age and gender.
    \end{tablenotes}
  \end{threeparttable}}
\end{table}

\begin{table}[H]
  \centering
  \caption{The Urban Wage Premium Holding Job Mobility Constant: Log Hourly Wages}
  \label{tab:decomposing_uwp_stayers_log}
  \resizebox{\linewidth}{!}{\begin{threeparttable}
    \begin{tabular}{l c c c c c}
      \toprule
      \toprule
      & (1) & (2) & (3) & (4) & (5) \\
      \cmidrule(lr){2-6}
      & Baseline (Full Sample) & Baseline (Stayers) & + Lagged Firm FE & 
      Coworker Effects & + Lagged Firm FE \\
      \cmidrule(lr){1-6} \\
      \multicolumn{6}{l}{\textbf{Panel A: First Stage Regressions}} \\
      \\
      Lagged Wage ($\nu$) & 0.8313 & 0.9526 & 0.8068 & 0.9768 & 0.9529 \\
                 & (0.0005) & (0.0004) & (0.0013) & (0.0003) & (0.0024) \\
            Higher-Wage Coworkers ($\tilde{\theta}^+_1$) & & & & 0.1376 & 0.1151 \\
                      & & & & (0.0024) & (0.0034) \\
            Lower-Wage Coworkers ($\tilde{\theta}^-_1$) & & & & 0.0238 & 0.0499 \\
                       & & & & (0.0013) & (0.0043) \\
            Nonlinear Effects ($\tilde{\theta}^{+}_2$) & & & & 0.0638 & 0.0686 \\
            & & & & (0.0016) & (0.0018) \\
            Nonlinear Effects ($\tilde{\theta}^{-}_2$) & & & & -0.0064 & 0.0131 \\
            & & & & (0.0008) & (0.0019) \\ \\
            Lagged Firm Fixed Effects & No & No & Yes & No & Yes \\
            \\
            Observations & 15,344,033 & 9,175,509 & 8,139,602 & 8,195,025 & 8,008,451 \\
            $R^2$ & 0.7531 & 0.8906 & 0.9042 & 0.8888 & 0.9076 \\
            \\
            \multicolumn{6}{l}{\textbf{Panel B: Second Stage Regressions}} \\
            \\
            Log Population ($\alpha$) & 0.00813 & 0.00302 & 0.00141 & 0.00125 & 0.00016 \\
                                 & (0.0008) & (0.0004) & (0.0005) & (0.0003) & (0.0004) \\
                            \\
                            Observations & 304 & 304 & 304 & 304 & 304 \\
                              $R^2$ & 0.3756 & 0.1757 & 0.0351 & 0.0324 & 0.0006 \\
                              \\
                              $\hat\alpha$ change relative to (1) & -- & -62.9\% & -82.7\% & -84.6\% & -98.0\% \\
                              $R^2$ change relative to (1) & -- & -53.2\% & -90.7\% & -91.4\% & -99.8\% \\     \bottomrule     \bottomrule
    \end{tabular}
    \begin{tablenotes}
      \footnotesize
      \item \textit{Note:} Panel A reports first-stage regressions where the dependent variable is log hourly wages. Column (1) shows the baseline specification from \Cref{eq:reg1} for the full sample. Columns (2)-(5) restrict to job stayers. Column (2) shows the baseline specification from \Cref{eq:reg1} restricting the sample to job stayers. Column (3) adds firm fixed effects following \Cref{eq:reg2}. Column (4) includes the full coworker effects specification from \Cref{eq:reg4}. Column (5) combines coworker effects with firm fixed effects following \Cref{eq:reg5}. Panel B reports the projection of estimated fixed effects from the first stage onto logged commuting zone population following \Cref{eq:FE_projection}. Standard errors are clustered at the commuting zone level. Wages are residualized for age and gender.
    \end{tablenotes}
  \end{threeparttable}}
\end{table}

\begin{table}[H]
  \centering
  \caption{Job-to-Job Transition Rates and City Size: Log Hourly Wages}
  \label{tab:EE_mechanisms_log}
  \begin{threeparttable}
    \begin{tabular}{l c c}
      \toprule 
      \toprule 
      & (1) & (2) \\
      \cmidrule(lr){2-3}
      Dependent variable & EE Transition & EE Transition \\
      \cmidrule(lr){1-3} \\
      
      Log Population & 0.0296 & 0.0468 \\
                    & (0.0017) & (0.0030) \\
      Lagged Log Wage & & -0.1032 \\
                    & & (0.0014) \\\\
      
      Controls & No & Lagged Wage \\
      Fixed Effects & No & No \\
      SE Clustering & CZ & CZ \\ \\
      
      Observations & 15,344,033 & 15,344,033 \\
      $R^2$ & 0.0023 &  0.0175 \\
      
      \bottomrule
      \bottomrule
    \end{tabular}
    \begin{tablenotes}
      \footnotesize
      \item \textit{Note:} This table reports regressions of job-to-job (EE) transition rates on city size. Column (1) shows the baseline specification from \Cref{eq:EE_1}. Column (2) adds lagged log hourly wages as a control following \Cref{eq:EE_2}. Standard errors are clustered at the commuting zone (CZ) level. The sample consists of all workers from one cross-section of the short panel.
    \end{tablenotes}
  \end{threeparttable}
\end{table}

\subsection{Robustness to Wage Growth}\label{app:wage_growth}
\begin{table}[H]
  \centering
  \caption{Decomposing the Urban Wage Premium: Wage Growth}
  \label{tab:decomposing_uwp_wage_growth}
  \resizebox{\linewidth}{!}{\begin{threeparttable}
    \begin{tabular}{l c c c c}
      \toprule 
      \toprule 
      & (1) & (2) & (3) & (4) \\
      \cmidrule(lr){2-5}
      & Baseline & + Lagged Firm FE & 
      Coworker Effects & + Lagged Firm FE \\
      \cmidrule(lr){1-5} \\
      \multicolumn{5}{l}{\textbf{Panel A: First Stage Regressions}} \\
      \\
           Higher-Wage Coworkers ($\tilde{\theta}^+_1$) & & & 0.0076 & 0.0155 \\
                            & & & (0.0001) & (0.0003) \\
                              Lower-Wage Coworkers ($\tilde{\theta}^-_1$) & & & 0.0025 & 0.0004 \\
                                     & & & (0.0001) & (0.0001) \\
                              Nonlinear Effects ($\tilde{\theta}^{+}_2$) & & & -0.0000 & -0.0000 \\
                              & & & (0.0000) & (0.0000) \\
                              Nonlinear Effects ($\tilde{\theta}^{-}_2$) & & & 0.0000 & 0.0000 \\
                              & & & (0.0000) & (0.0000) \\ \\
                              Lagged Firm Fixed Effects & No & Yes & No & Yes \\
                              \\
                              Observations & 15,344,033 & 13,749,982 & 13,505,876 & 13,505,749 \\
                              $R^2$ & 0.0123 & 0.1462 & 0.0453 & 0.1953 \\
                              \\
                              \multicolumn{5}{l}{\textbf{Panel B: Second Stage Regressions}} \\
                              \\
                              Log Population ($\alpha$) & 0.00128 & 0.00135 &0.00054  & 0.00084 \\
                                   & (0.0003) & (0.0005) & (0.0003) & (0.0005) \\
                              \\
                              Observations & 304 & 304 & 304 & 304 \\
                              $R^2$ & 0.0594 & 0.0327 & 0.0098 & 0.0122 \\
                              \\
                              $\hat\alpha$ change relative to (1) & -- & +5.2\% & -58.0\% &  -34.2\%\\
                              $R^2$ change relative to (1) & -- & -45.0\% & -84.8\% & -79.5\% \\
      
      \bottomrule
      \bottomrule
    \end{tabular}
    \begin{tablenotes}
      \footnotesize
      \item \textit{Note:} Panel A reports first-stage regressions where the dependent variable is hourly wage growth. Column (1) shows the baseline specification from \Cref{eq:reg1'} with commuting zone fixed effects. Similarly, columns (2)-(4) report the results from the specifications where we change \Cref{eq:reg2,eq:reg4,eq:reg5} to use wage growth as the dependent variable. Panel B reports the projection of estimated fixed effects from the first stage onto logged commuting zone population following \Cref{eq:FE_projection}. Standard errors are clustered at the commuting zone level. Sample consists of all workers in the short panel with residualized wages for age and gender.
    \end{tablenotes}
  \end{threeparttable}}
\end{table}

\begin{table}[H]
  \centering
  \caption{Job Stayers with Wage Growth as Dependent Variable}
  \label{tab:decomposing_uwp_stayers_wage_growth}
  \resizebox{\linewidth}{!}{\begin{threeparttable}
    \begin{tabular}{l c c c c c}
      \toprule
      \toprule
      & (1) & (2) & (3) & (4) & (5) \\
      \cmidrule(lr){2-6}
      & Baseline (Full Sample) & Baseline (Stayers) & + Lagged Firm FE & 
      Coworker Effects & + Lagged Firm FE \\
      \cmidrule(lr){1-6} \\
      \multicolumn{6}{l}{\textbf{Panel A: First Stage Regressions}} \\
      \\
          Higher-Wage Coworkers ($\tilde{\theta}^+_1$) & & & & 0.0390 & 0.0083 \\
                      & & & & (0.0008) & (0.0002) \\
                Lower-Wage Coworkers ($\tilde{\theta}^-_1$) & & & & 0.0010 & -0.0001 \\
                       & & & & (0.0000) & (0.0000) \\
                Nonlinear Effects ($\tilde{\theta}^{+}_2$) & & & & -0.0000 & -0.0000 \\
                    & & & & (0.0000) & (0.0000) \\
                Nonlinear Effects ($\tilde{\theta}^{-}_2$) & & & & 0.0000 & -0.0000 \\
                    & & & & (0.0000) & (0.0000) \\ \\
                Lagged Firm Fixed Effects & No & No & Yes & No & Yes \\
                    \\
                Observations & 15,344,033 & 9,175,509 & 8,139,602 & 8,195,025 & 8,008,451 \\
                $R^2$ & 0.0123 & 0.0081 & 0.1886 & 0.0266 & 0.2154 \\
                    \\
                \multicolumn{6}{l}{\textbf{Panel B: Second Stage Regressions}} \\
                    \\
                Log Population ($\alpha$) & 0.00128 & 0.0007805 & 0.00003 & 0.0002909 & -0.0003749 \\
                       & (0.0003) & (0.000279) & (0.0002985) & (0.0003099) & (0.0003282) \\
                    \\
                Observations & 304 & 304 & 304 & 304 & 304 \\
                $R^2$ & 0.0594 & 0.0194 & 0.0000 & 0.0021 & 0.0043 \\
                    \\
                $\hat\alpha$ change relative to (1) & -- & -39.0\% & -97.7\% & -77.3\% & -129.3\% \\
                $R^2$ change relative to (1) & -- & -67.3\% & -100.0\% & -96.5\% & -92.8\% \\
      
      \bottomrule
      \bottomrule
    \end{tabular}
    \begin{tablenotes}
      \footnotesize
      \item \textit{Note:} Panel A reports first-stage regressions where the dependent variable is hourly wage growth. Column (1) shows the baseline specification from \Cref{eq:reg1'} with commuting zone fixed effects. Columns (2)-(5) restrict to job stayers. Column (2) shows the baseline specification from \Cref{eq:reg1'} restricting the sample to job stayers. Similarly, columns (2)-(4) report the results from the specifications where we change \Cref{eq:reg2,eq:reg4,eq:reg5} to use wage growth as the dependent variable. Panel B reports the projection of estimated fixed effects from the first stage onto logged commuting zone population following \Cref{eq:FE_projection}. Standard errors are clustered at the commuting zone level. Wages are residualized for age and gender.
    \end{tablenotes}
  \end{threeparttable}}
\end{table}

\subsection{Other Moments}\label{app:other_moments}

\begin{table}[H]
  \centering
  \caption{Adding Moments of the Coworker Wage Distribution}
  \label{tab:decomposing_uwp_moments}
  \resizebox{\linewidth}{!}{\begin{threeparttable}
    \begin{tabular}{l c c c c c c c c c}
      \toprule
        \toprule
        & (1) & (2) & (3) & (4) & (5) & (6) & (7) & (8) &(9)\\
        \cmidrule(lr){2-10}
        & Baseline & +Mean &  +Variance & +Skewness & +Kurtosis & +P1 & +P10 & +P90 & +P99 \\
         \multicolumn{10}{l}{\textbf{Panel A: First Stage Regressions}} \\
          \\
          Lagged Wage & 0.8072 & 1.0260 & 0.9649 & 0.8916 & 0.8968 & 0.7947 & 0.7140 & 0.6802 & 0.6900 \\
           & (0.0442) & (0.0913) & (0.0920) & (0.1068) & (0.1126) & (0.1081) & (0.0858) & (0.0773) & (0.0757) \\
          Mean Wage &  & -0.1713 & -0.1007 & -0.0837 & -0.0857 & -0.1688 & -0.3586 & -0.4858 & -0.5065 \\
            &  & (0.1356) & (0.1397) & (0.1466) & (0.1474) & (0.1539) & (0.1729) & (0.1529) & (0.1541) \\
          Mean Wage $\times$ Lagged Wage &  & -0.0009 & -0.0012 & -0.0011 & -0.0011 & -0.0011 & -0.0009 & -0.0007 & -0.0007 \\
            &  & (0.0002) & (0.0003) & (0.0003) & (0.0003) & (0.0003) & (0.0003) & (0.0001) & (0.0001) \\
          Variance &  &  & 0.0000 & 0.0000 & 0.0000 & 0.0000 & 0.0000 & -0.0000 & -0.0000 \\
           &  &  & (0.0000) & (0.0000) & (0.0000) & (0.0000) & (0.0000) & (0.0000) & (0.0000) \\
          Variance $\times$ Lagged Wage &  &  & 0.0000 & 0.0000 & 0.0000 & 0.0000 & 0.0000 & 0.0000 & 0.0000 \\
           &  &  & (0.0000) & (0.0000) & (0.0000) & (0.0000) & (0.0000) & (0.0000) & (0.0000) \\
          Skewness &  &  &  & -0.1129 & 0.1150 & -0.2209 & -0.5306 & -0.6622 & -0.7848 \\
           &  &  &  & (0.1210) & (0.2096) & (0.1748) & (0.1662) & (0.1501) & (0.2140) \\
          Skewness $\times$ Lagged Wage &  &  &  & 0.0150 & 0.0121 & 0.0253 & 0.0315 & 0.0349 & 0.0344 \\
           &  &  &  & (0.0043) & (0.0082) & (0.0075) & (0.0084) & (0.0074) & (0.0095) \\
          Kurtosis &  &  &  &  & -0.0099 & -0.0021 & 0.0003 & 0.0063 & 0.0085 \\
           &  &  &  &  & (0.0040) & (0.0032) & (0.0030) & (0.0031) & (0.0048) \\
          Kurtosis $\times$ Lagged Wage &  &  &  &  & 0.0001 & -0.0000 & -0.0003 & -0.0003 & -0.0003 \\
           &  &  &  &  & (0.0001) & (0.0001) & (0.0001) & (0.0001) & (0.0002) \\
          P1 &  &  &  &  &  & 0.3272 & -0.1359 & -0.0535 & -0.0232 \\
          &  &  &  &  &  & (0.0768) & (0.0944) & (0.0859) & (0.1004) \\
          P1 $\times$ Lagged Wage &  &  &  &  &  & 0.0019 & 0.0008 & -0.0008 & -0.0011 \\
          &  &  &  &  &  & (0.0005) & (0.0035) & (0.0031) & (0.0035) \\
          P10 &  &  &  &  &  &  & 0.9029 & 0.6541 & 0.6599 \\
             &  &  &  &  &  &  & (0.2190) & (0.1827) & (0.1844) \\
          P10 $\times$ Lagged Wage &  &  &  &  &  &  & 0.0007 & 0.0024 & 0.0024 \\
             &  &  &  &  &  &  & (0.0032) & (0.0035) & (0.0036) \\
          P90 &  &  &  &  &  &  &  & 0.1919 & 0.1491 \\
          &  &  &  &  &  &  &  & (0.0380) & (0.0323) \\
          P90 $\times$ Lagged Wage &  &  &  &  &  &  &  & -0.0004 & -0.0001 \\
          &  &  &  &  &  &  &  & (0.0006) & (0.0003) \\
          P99 &  &  &  &  &  &  &  &  & 0.0218 \\
             &  &  &  &  &  &  &  &  & (0.0074) \\
          P99 $\times$ Lagged Wage &  &  &  &  &  &  &  &  & -0.0001 \\
             &  &  &  &  &  &  &  &  & (0.0001) \\
          \\
          Observations & 15,344,033 & 13,505,876 & 13,149,735 & 13,139,708 & 13,139,708 & 13,139,708 & 13,139,708 & 13,139,708 & 13,139,708 \\
          $R^2$ & 0.1092 & 0.1098 & 0.1270 & 0.1280 & 0.1280 & 0.1314 & 0.1360 & 0.1619 & 0.1628 \\
          \\
          \multicolumn{10}{l}{\textbf{Panel B: Second Stage Regressions}} \\
          \\
          Log Population ($\alpha$) & 0.23614 & 0.2413 & 0.24697 & 0.24220 & 0.23448 & 0.26442 & 0.15896 & 0.12458 & 0.13569 \\
               & (0.0278) & (0.0301) & (0.0296) & (0.0277) & (0.0277) & (0.0304) & (0.0236) & (0.0127) & (0.0121) \\
          \\
          $\hat\alpha$ change relative to (1) & -- & +2.2\% & +4.6\% & +2.6\% & -0.7\% & +12.0\% & -32.7\% & -47.2\% & -42.5\% \\
          $R^2$ change relative to (1) & -- & -5.2\% & -4.0\% & -6.2\% & -9.4\% & -25.3\% & -17.6\% & -1.8\% & +2.0\% \\
          \bottomrule
        \bottomrule
    \end{tabular}
    \begin{tablenotes}
      \footnotesize
      \item \textit{Note:} Panel A reports first-stage regressions where the dependent variable is hourly wages. All specifications include commuting zone fixed effects, age, and gender controls. Column (1) includes lagged wages. Column (2) adds mean wage and its interaction with lagged wage. Column (3) adds variance and its interaction with lagged wage. Column (4) adds skewness and its interaction. Column (5) adds kurtosis and its interaction. Column (6) adds the 1st percentile (P1) and its interaction. Column (7) adds the 10th percentile (P10) and its interaction. Column (8) adds the 90th percentile (P90) and its interaction. Column (9) adds the 99th percentile (P99) and its interaction. Panel B reports the projection of estimated commuting zone fixed effects from the first stage onto logged commuting zone population. Robust standard errors reported. Sample consists of all workers in the short panel with residualized wages for age and gender.
    \end{tablenotes}
  \end{threeparttable}}
\end{table}

\end{document}




